# Band Gap Engineering of Two-Dimensional Nitrogene

Jie-Sen Li[1,2], Wei-Liang Wang[1], Dao-Xin Yao[1]*

In our previous study, we have predicted the novel two-dimensional honeycomb monolayers of pnictogen. In particular, the structure and properties of the honeycomb monolayer of nitrogen, which we call nitrogene, are very unusual. In this paper, we make an in-depth investigation of its electronic structure. We find that the band structure of nitrogene can be engineered in several ways: controlling the stacking of monolayers, application of biaxial tensile strain, and application of perpendicular electric field. The band gap of nitrogene is found to decrease with the increasing number of layers. The perpendicular electric field can also reduce the band gap when it is larger than 0.18V/ Å, and the gap closes at 0.35V/ Å. A nearly linear dependence of the gap on the electric field is found during the process. Application of biaxial strain can decrease the band gap as well, and eventually closes the gap. After the gap-closing, we find six inequivalent Dirac points in the Brillouin zone under the strain between 17% and 28%, and the nitrogene monolayer becomes a Dirac semimetal. These findings suggest that the electronic structure of nitrogene can be modified by several techniques, which makes it a promising candidate for electronic devices.

[1] State Key Laboratory of Optoelectronic Materials and Technologies, School of Physics, Sun Yat-Sen University, Guangzhou, P. R. China.
[2] School of Environment and Chemical Engineering, Foshan University, Foshan, P. R. China
Correspondence and requests for materials should be addressed to D. X. Y. (email: yaodaox@mail.sysu.edu.cn)

Since the discovery of graphene [1], a two-dimensional (2D) carbon-based material, the study on the two-dimensional materials has been hot topic for scientific endeavor. The applications of materials have strong dependence on its mechanical, thermal, and electric properties, especially in the field of electronic devices [2]. However, the lack of intrinsic band gap of graphene imposes the restriction on its applications.

Recently various 2D materials have been predicted and synthesized successfully. For example, silicene and germanene, and cousins of graphene have been found and show interesting electronic structures[3]. In these materials, the strong spin-orbit couplings can open energy gap in the Brillouin zone, which gives silicene and germanene an advantage over graphene in real applications. In 2015, stanene was also predicted, and its topological aspects have been found[4].

In early 2014, D. Tomanak's group predicted the existence of 2D blue phosphorene, a new phase of phosphorus[5]. Several months later, they predicted two new phases of 2D structure of phosphorus[6]. In 2015, we proposed the 2D honeycomb monolayer of pnictogen [7]. Especially, we predicted the existence of 2D nitrogen honeycomb monolayer, which we call "nitrogene". However, there is still no in-depth investigation of the electronic structure of nitrogene under various conditions. In particular, the control of its electronic structure has not been studied.

There are mainly three viable ways of controlling the electronic structure: 1) change its composition and structure, such as tuning the band gap of graphene by adjusting the degree of hydrogenation[8]; 2) apply an external field, such as the perpendicular electric field on arsenene[9] and bilayer graphene[10]; 3) use a strain, which has been proven to be applicable for most of 2D materials[9]. Since atoms in nitrogene have achieved octet stability, it is very difficult to form bonds with other atoms or functional groups. Therefore, the controlling of the electronic structure of nitrogene may be restricted to external field and strain. In this paper, we use first-principle calculation to investigate the band structure of the newly predicted nitrogene under various conditions. We show that nitrogene may have promising applications, such as ultra-capacitors[11], field effect transistors (FETs)[12,13], bio and chemical sensors[14,15].

# Results

**Structure and Stability**

Nitrogen gas is the most stable form of nitrogen, and all other allotropes may not be stable under normal conditions. In order to investigate the stability of nitrogene, we calculate its phonon dispersion using first-principle techniques. As shown in Figure 1, no vibration modes with imaginary frequency is found for the whole Brillouin zone, which suggests its stability. Based on this dispersion, the sound speed of three acoustic branches can be obtained. For the two transverse modes, the sound speeds are 317m/s and 1600m/s, respectively. The speed of the transverse mode with lower frequency is very close to the speed of sound wave in the air. The acoustic longitudinal mode has a sound speed of 2524m/s, which is lower than most of the 2D materials. The sound speed of the acoustic longitudinal mode is closely related to the in-plane rigidity. The result of phonon dispersion of nitrogene shows that nitrogene has moderate Young's modulus, and therefore has great tolerance for lattice mismatch when attached to other materials. That means it is possible to synthesis nitrogene on various substrates.

**Multi-Layer Nitrogene**

In our previous study, single layer nitrogene has been shown to be a semiconductor with indirect band gap of 3.7eV[7]. For three-dimensional nitrogene, the band gap decreases to 2.1eV, and the band structures of both single layer nitrogene and 3D nitrogene are shown in Figure 2. As the number of layers increases, the band gap decreases due to the inter-layer coupling, as shown in Figure 3. The band gap dependence on number of layers for the multi-layer nitrogene can be approximated by the following empirical formula:

$$E_g = \frac{2.3}{n} + 1.9 (\text{eV}) \quad (n \geq 2) \tag{1}$$

where $n$ is the number of layers. Similar to its 2D counterpart, the band structure of 3D nitrogene can be divided into three branches. Compared with the 2D band structure, each the band in 3D nitrogene splits into two bands, but the splitting is relatively small, and some of these bands are even degenerate. This is an indication that the couplings are weak van de Waals interactions. For the band structure of multilayer nitrogene, each band splits into several bands, depending on the number of layers. And some of these bands are still degenerate, as shown in Figure 4. The band structure of multilayer nitrogene can be regarded as a projection of 3D band structure onto its 2D counterpart. For all these cases, the component of 2p orbitals dominates the middle branch and upper branch, while the lower branch has about the same contribution from 2p orbitals and 2s orbitals, as shown in the density of states (DOS) in Figure 2. In both the multilayer case and 3D case, van de Waals interaction broadens the band width and reduces the band gap.

**The Effect of Electric Field**

As mentioned before, the application of electric field is a useful technique to control the band structure of materials. Since graphene is planar, the application of perpendicular electric field has no obvious effect on its gap. In case of nitrogene, the band gap depends on the perpendicular electric field. When no electric field is present, the bands of nitrogene show three branches, which are represented in three different colors in Figure 2. For the sake of the following discussion, we refer to these eight bands as intrinsic bands. Based on our first-principle calculation, the intrinsic bands are not affected by the electric field. However, the high energy bands move down due to the electric field. When the electric field reaches 0.18V/ Å, the conduction band minimum shifts to Γ point, and the gap starts to decrease if the electric field continues to increase. The band gap closes when the electric field reaches 0.35V/ Å, as shown in Figure 5(a-d). Similar results can be found in arsenene[9]. In the interval of 0.18V/ Å and 0.35V/ Å, the band gap has linear dependence on electric field, as shown in Figure 5(e). This phenomenon can be attributed to the giant Stark effect, which arises from the redistribution of electron density in both the valance band maximum (VBM) and the conduction band minimum (CBM)[16,17]. Similar effect can be found in other systems, such as double-stranded porphyrin ladder polymers[18], phosphorene[19], and boron-carbon-nitride[20].

**The Effect of Biaxial Tensile Strain**

The structural parameters of nitrogene change with the strain. Structural optimization shows that nitrogene decomposes into nitrogen gas under the 23% strain. However, the meta-stable state of nitrogene can exist when the strain is more than 23%, as long as the external perturbation is small enough. Nitrogene retains its buckled structure when the strain is between 23% and 27%. When the strain reaches 28% and the bucking distance becomes zero, which means it becomes planar. As the strain increases, the buckling distance increases at the beginning, and reaches a maximum of 1.82 Å under 17% strain. The buckling distance starts to decrease after the strain surpasses 17%, and it suddenly becomes zero as strain reaches 29%. The initial increase of buckling distance may have to do with the weakening of N-N bond: as the bond angle increases, the $\sigma$ bonds are bent, which reduces the overlap between the atomic orbitals. From 0 to 28% strain, the bond length increases monotonously The bond length drops a little due to the transition from the buckled structure to the planar structure at 29% strain, and it continues to increases after 29% strain. They have been summarized in Figure 6.

As the strain increases, the original indirect band gap changes to a direct one, and the gap continues to decrease, which can be seen in Figures 7(a) and (b). The three bands below the Fermi level come from the bonding orbitals while the three bands above the Fermi level are from the anti-bonding orbitals. The increase of the lattice constant decreases the buckling distance, which means the bonding angles increase, and the p character in the hybridizations of atoms decreases, which makes the orbital overlap smaller. On the other hand, the increased distance also weakens the bonds. Therefore, the energy difference between the bonding orbitals and the anti-bonding orbitals decreases, and reduces the band gap. The gap closes when the strain is ~17%, and the system becomes metallic hereinafter, as shown in Figure 8. When the gap closes, the conduction band minimum and the valence band maximum touch each other at the Γ point, as shown in Figure 7(c) and Figure 9. As the strain continues to increase, the conduction band moves below the Fermi level while the valence band is above it, therefore these two bands intersect with each other, as shown in Figure 7 (d)-(e), and Figure 9. A gap opens on the line connecting the Γ point to M point in the Brillouin zone, and the line connecting the Γ point to K point remains gapless. A Dirac cone appears between the Γ point to K point, as shown in Figure 7(d)-(e), and Figure 9. A linear dispersion near the Fermi level is found in Figure 10. These Dirac cones are anisotropic, *i.e.* the Fermi velocity is angle-dependent. Under the strain between 16% to 28%, there are six Dirac points in the Brillouin zone, and the system is metallic. Under the 28% strain, the six Dirac points are located at $\frac{0.53\pi}{a}\left(\frac{\sqrt{3}}{2},\pm\frac{1}{2}\right)$, $\frac{0.53\pi}{a}\left(-\frac{\sqrt{3}}{2},\pm\frac{1}{2}\right)$, and $\frac{0.53\pi}{a}(0,\pm 1)$, respectively. Under this strain, the distance from the Dirac points to the Γ point is $\frac{0.53\pi}{a}$, which is about 40% of $\frac{4\pi}{3a}$, the distance from the K points to the Γ point.

When the strain reaches 30%, the band structure becomes dramatically different because of the sudden change of structure. In addition to the Dirac point along the Γ-K line, another Dirac point appears between the M point and K point, as shown in Figure 7(f), and the former is higher in energy than the latter. After the structure transition, the Fermi energy no longer lies in the original Dirac points, but lies in the new Dirac points instead. The original Dirac points are above the Fermi energy, and the system changes from a Dirac semimetal to a metal. Interestingly, after the gap closes, the band above the Fermi level becomes very flat along the Γ-M line in the BZ, which is responsible for the sharp peak near the Fermi energy in the density of states (Figure 9). The M point in this band is a saddle point, which gives rise to a Van Hove singularity. This Van Hove singularity near the Fermi level may be closely related to a chiral p-wave superconductivity[21].

**Discussion**

In this paper, we verify the stability of nitrogene, and show how the electronic structure of nitrogene can be tuned by various techniques. The calculated phonon spectrum of nitrogene shows the stability of nitrogene, and the soft longitudinal mode suggests the great tolerance of lattice mismatch and the possible synthesis of nitrogene on substrates. The electronic structure of multilayer nitrogene is investigated, and the gap decreases with increasing the number of layers. The splitting of the band structures of multi-layer nitrogene shows the existence of van de Waals interlayer interaction. Perpendicular electric field shows no effect to the intrinsic band structure of nitrogene, but it can lower the energy of the bands far above the Fermi energy, and thus changes the band gap. We find that the gap decreases linearly with respect to the perpendicular electric field in the interval of 0.18V/ Å and 0.35V/ Å. This can be explained by giant stark effect that arises from the formation of dipole inside the materials. The application of strain can also reduce the band gap, and the gap closes under the 17% strain. Interestingly, the strain between 17% and 28% can give rise to six inequivalent Dirac points in the Brillouin zone. We find that the Fermi level lies on the Dirac points, therefore, the system becomes Dirac semimetal. The Fermi velocities on these Dirac points are angle-dependent, which makes it very different from graphene. The linear dispersion means that the electrons near Fermi level behaves as massless Dirac fermions, and the material may have novel transport properties under the strain. The Van Hove singularity under the strain makes it a candidate for the topological superconductor. All these findings show that nitrogene would have promising applications in electronics, optics, mechanics, and solar cells.

# Methods

Our calculations are based on Plane Augmented Wave (PAW) with Perdew-Burke-Ernzerh (PBE) of exchange-correlation as implemented in the Vienna Ab initio Simulation Package (VASP) code[22]. The systems are restricted to periodic boundary conditions. A vacuum at least 15 Å thick is inserted to eliminate the interaction between different layers. For optimization, ions are relaxed using conjugate-gradient algorithm until the total force is less than 0.01eV/ Å. The Brillouin zone is sampled by 8×8×1 grid with Monkhorst-Pack scheme for band structure under perpendicular electric, and 20×20×1 grid with Monkhorst-Pack scheme for other calculations. The phonon dispersion is calculated by interface of Phonopy to VASP[23]. A 4×4×1 supercell is used for the calculation of phonon. Since the spin-orbit coupling (SOC) in nitrogene is negligibly small, as we mentioned in our previous study[7], we do not include SOC in our calculations.

# Acknowledgements

The authors thank F. C. Chuang and W. X. Lin for helpful discussions. This project is supported by NBRPC-2012CB821400, NSFC-11574404, NSFC-11275279, NSFC-11274393, Natural Science Foundation of Guangdong Province (2015A030313176), Special Program for Applied Research on Super Computation of the NSFC-Guangdong Joint Fund, and Fundamental Research Funds for the Central Universities of China.

# Author contributions

J. S. L. performed of all the first-principle calculations, and analyzed the results of these calculations. W. L. W. was in charge of commercial software usage. D. X. Y. was the leader of this study. J. S. L. and D. X. Y. wrote the paper and all authors commented on it.

# Additional information

Competing financial interests: The authors declare no competing financial interests.

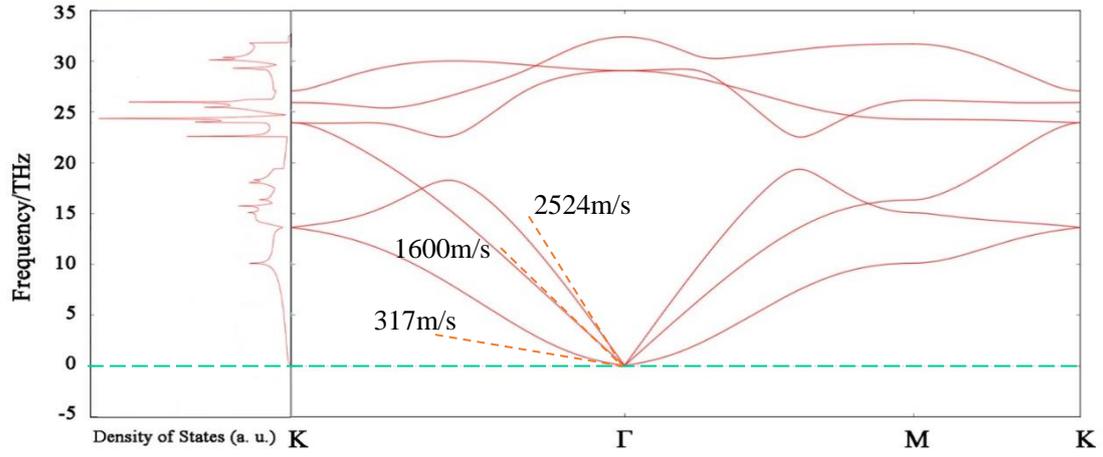

Figure 1 Phonon dispersion of nitrogene monolayer (right panel) and the corresponding density of states (left panel).

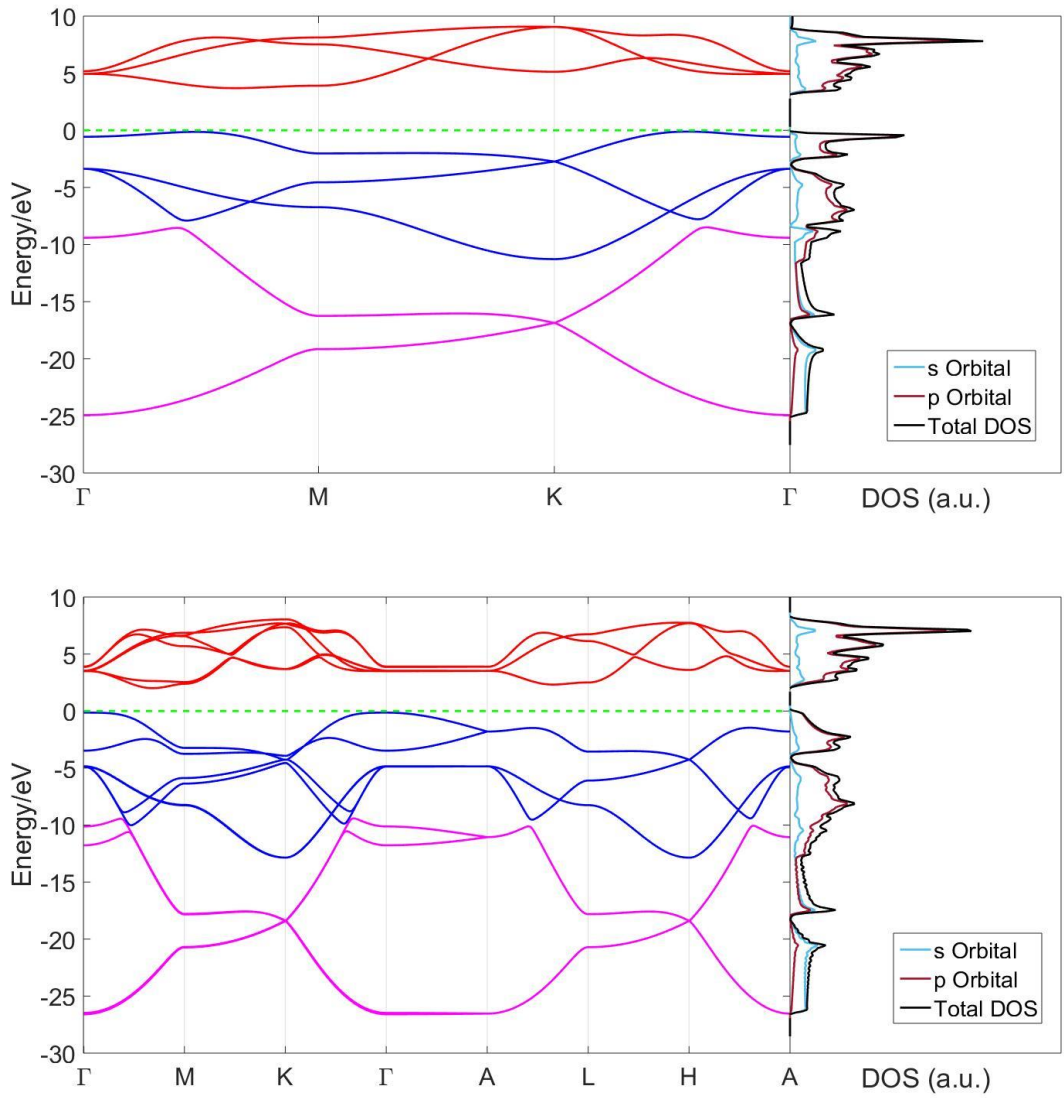

Figure 2 Comparison of the band structure (left panel) and density of states (right panel) of single layer nitrogene (upper panel) and 3D nitrogene (lower panel).

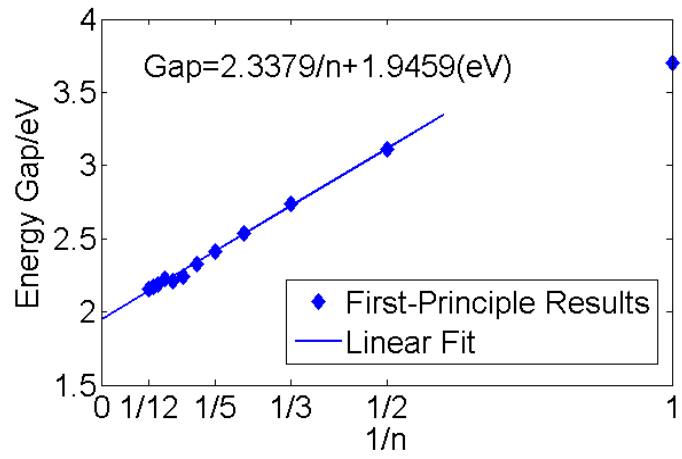

Figure 3 Dependence of band gap on the number of layers.

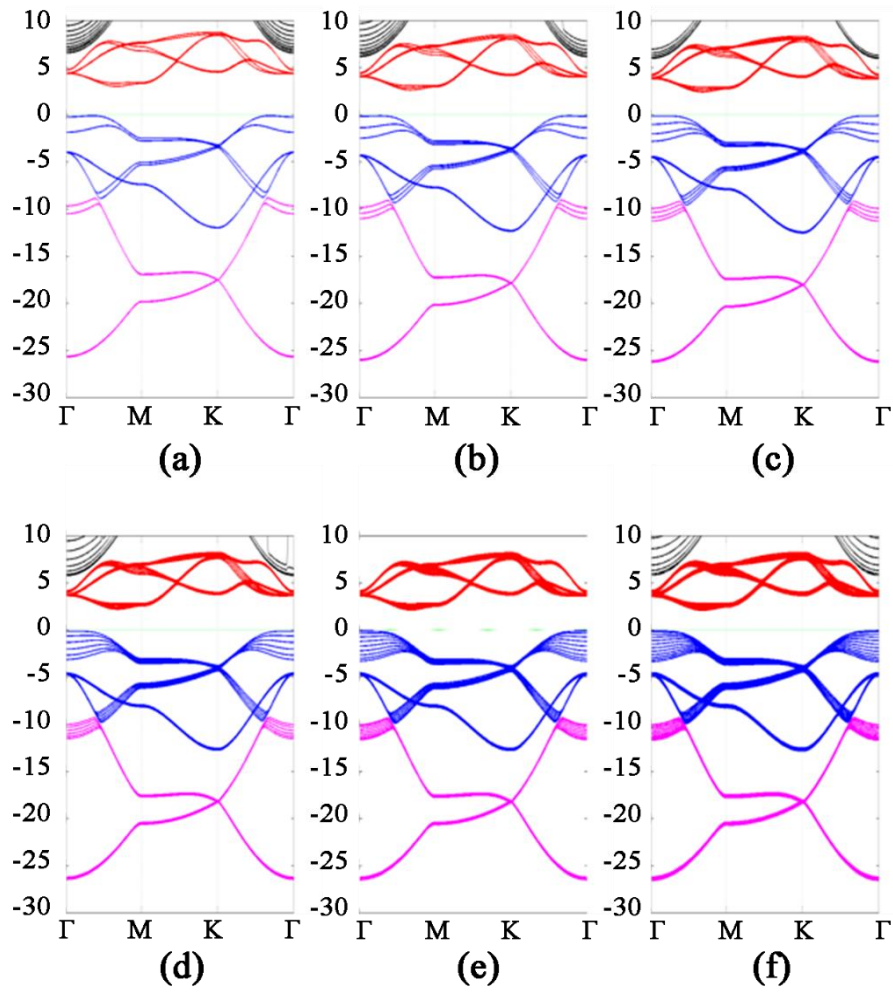

Figure 4 Band structure of multilayer nitrogene with layer number (a) two, (b) three, (c) four, (d) six, (e) nine, and (f) twelve, respectively.

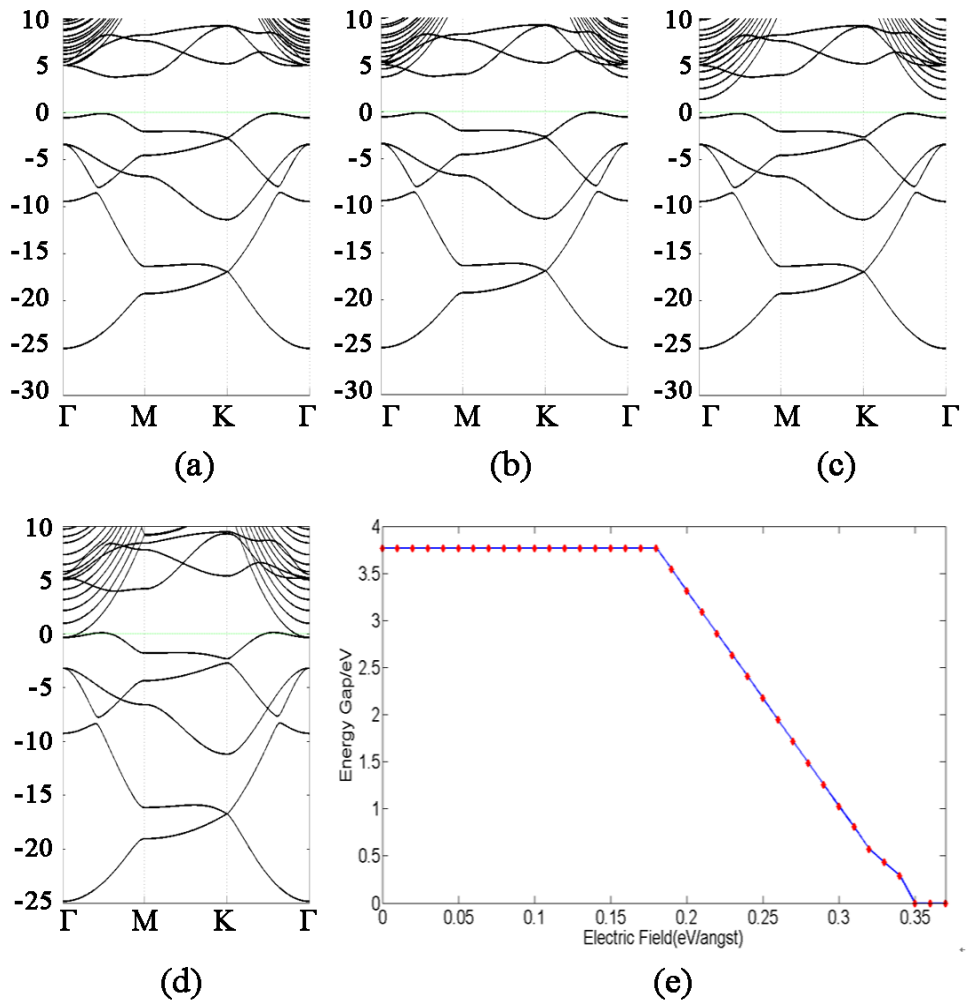

Figure 5 Band structures of nitrogene under perpendicular electric field of, (a) 0.12eV/ Å, (b) 0.18eV/ Å, (c) 0.28eV/ Å, and (d) 0.35eV/ Å, respectively. (e) Dependence of energy gap on the perpendicular electric field.

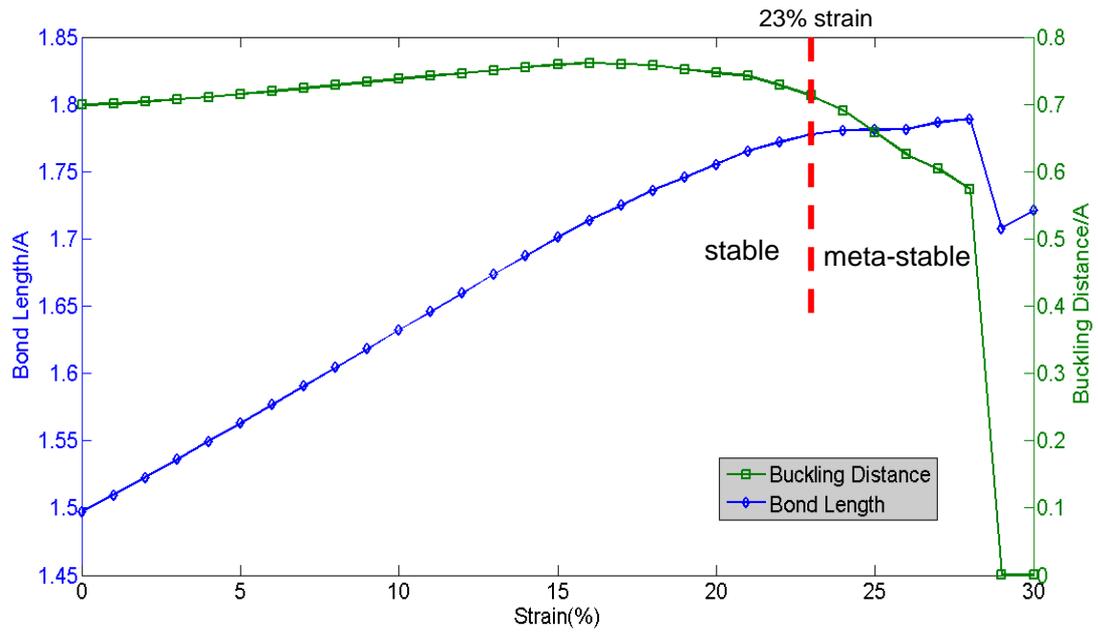

Figure 6 Dependence of covalent bond length (blue) and buckling distance (green) on strain.

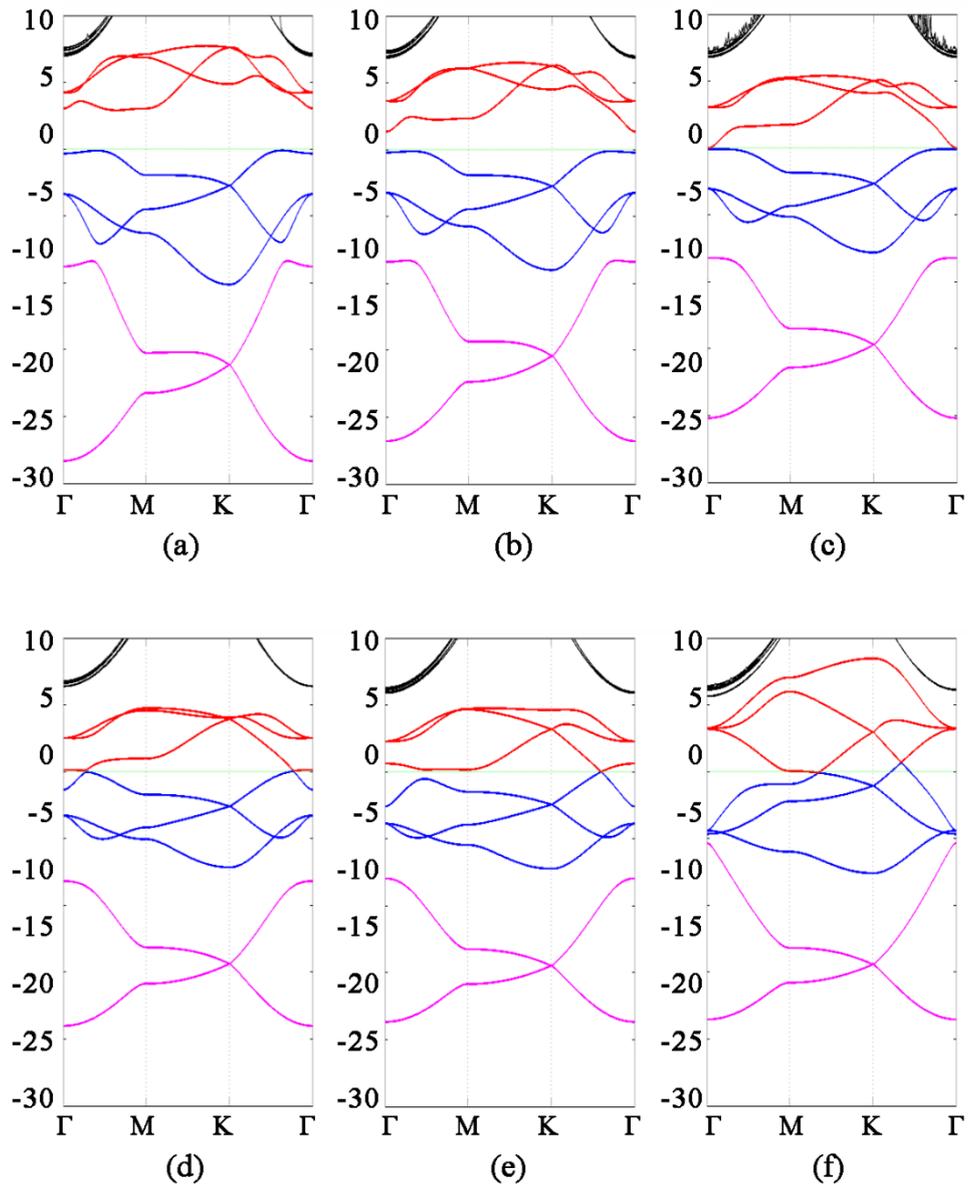

Figure 7 Band structure of nitrogene under (a)5%, (b)10%, (c)17%, (d)23%, (e)28%, and (f)30% strain

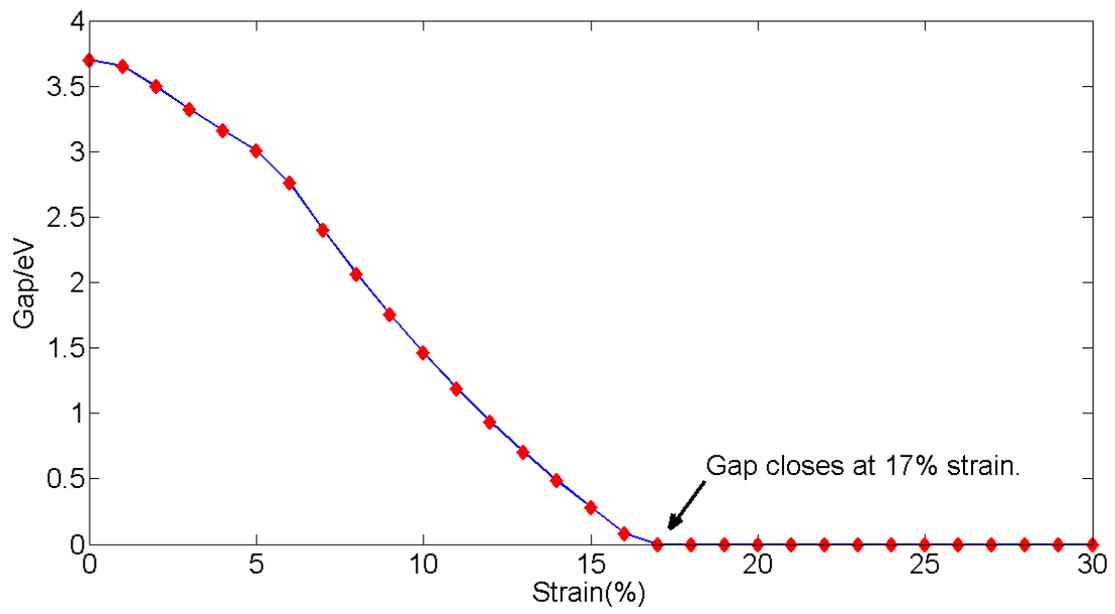

Figure 8 Dependence of energy gap on strain

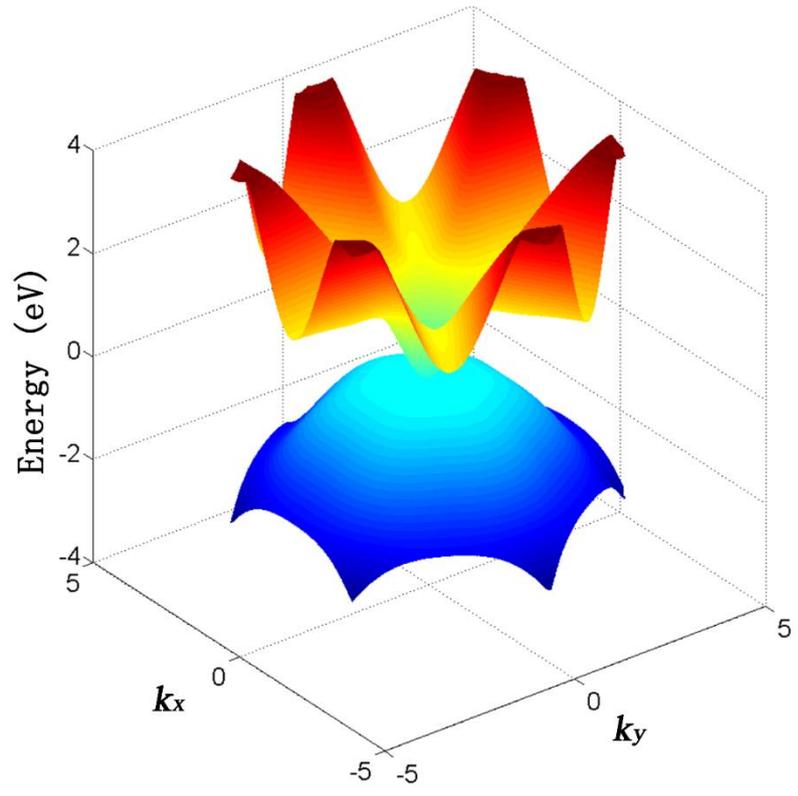

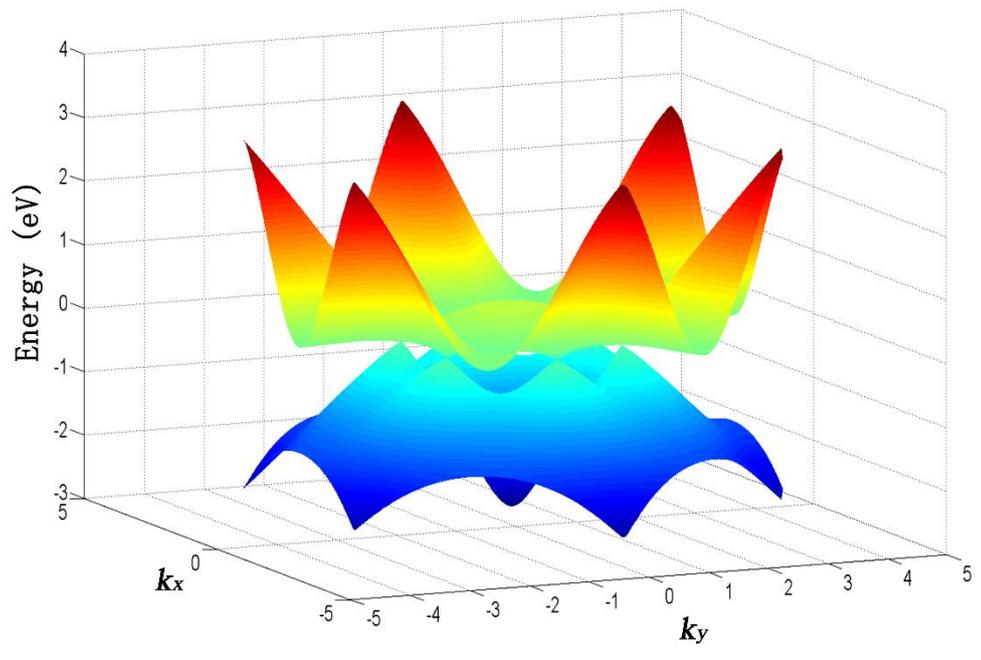

Figure 9 Two energy bands of nitrogene under (upper panel) 17% and (lower panel) 28% strains near the Fermi energy

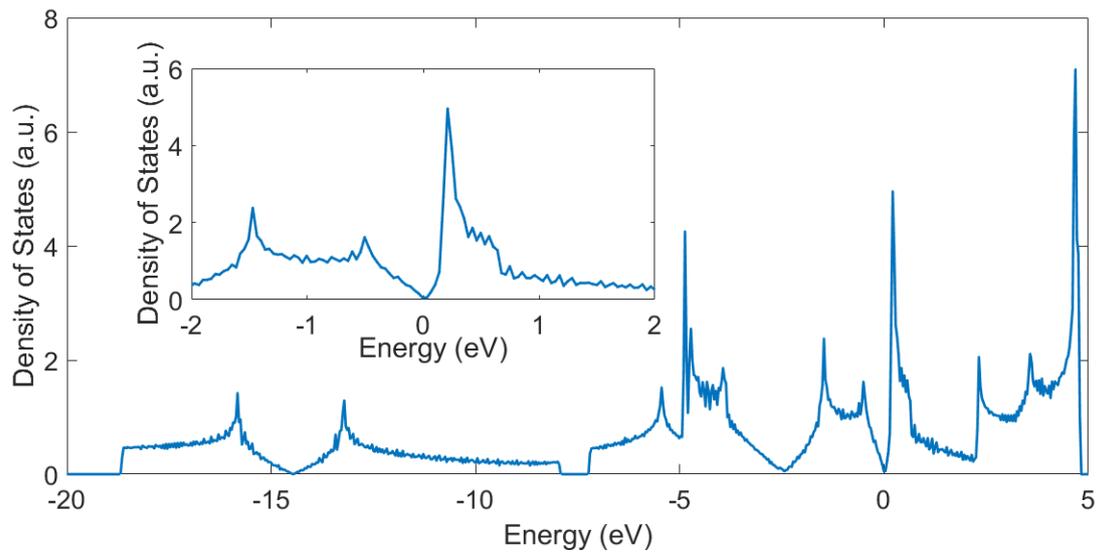

Figure 10 Density of states of nitrogene under 28% strain. Inset: Density of states from -2eV to 2eV.